\documentstyle[epsfig,12pt]{article}
\def \bea{\begin{eqnarray}}
\def \beq{\begin{equation}}

\def \eea{\end{eqnarray}}
\def \eeq{\end{equation}}
\def \gto{\stackrel{\gamma}{\to}}

\def \ups{\Upsilon}
\begin{document}
\Large
\centerline {\bf Production of singlet P-wave $c\bar{c}$ and $b \bar
b$ states\footnote{Enrico Fermi Institute preprint EFI 02-79, 
hep-ph/0205255.
Submitted to Physical Review D.}}
\normalsize
\bigskip
 
\centerline{Stephen Godfrey~\footnote{godfrey@physics.carleton.ca}}
\centerline{\it Ottawa-Carleton Institute for Physics}
\centerline{\it Department of Physics, Carleton University}
\centerline{\it 1125 Colonel By Drive, Ottawa, ON K1S 5B6, Canada}
\smallskip
\centerline{and}
\smallskip
\centerline{Jonathan L. Rosner~\footnote{rosner@hep.uchicago.edu}}
\centerline {\it Enrico Fermi Institute and Department of Physics}
\centerline{\it University of Chicago, 5640 S. Ellis Avenue, Chicago, IL 60637}
\bigskip
 
\begin{quote}

No spin-singlet $b \bar b$ quarkonium state has yet been observed.  In this
paper we discuss the production of the singlet P-wave $b\bar{b}$ and $c\bar{c}$
$^1P_1$ states $h_b$ and $h_c$.  We consider two possibilities.  In the first
the $^1P_1$ states are produced via the electromagnetic cascades $\ups(3S) \to
\eta_b(2S) + \gamma  \to h_b + \gamma \gamma \to  \eta_b  +\gamma\gamma\gamma$
and $\psi'\to \eta_c' + \gamma \to h_c + \gamma \gamma \to \eta_c +
\gamma\gamma\gamma$.  A more promising process consists of single pion
transition to the $^1P_1$ state followed by the radiative transition to the
$1^1S_0$ state:  $\ups(3S)\to h_b + \pi^0 \to \eta_b + \pi^0 +\gamma$ and
$\psi' \to h_c + \pi^0 \to \eta_c + \pi^0 +\gamma$.  For a million
$\ups(3S)$ or $\psi'$'s produced we expect these processes to produce
several hundred events.

\end{quote}
\bigskip

\noindent
PACS Categories:  14.40.Gx, 13.20.Gd, 13.40.Hq, 12.39.Ki

\bigskip

The study of bound states of heavy quarks has provided important tests of 
quantum chromodynamics (QCD) \cite{revs}.  The heavy quarkonium $c\bar{c}$ and
$b\bar{b}$ resonances have a rich spectroscopy with numerous narrow $S$, $P$,
and $D$-wave levels below the production threshold of open charm and beauty
mesons.  The spin-triplet $S$-wave states, $\psi(nS)$ and 
$\ups(nS)$ with $J^{PC}=1^{--}$, are readily produced by virtual 
photons in $e^+e^-$ or hadronic interactions, and then undergo 
electric dipole (E1) transition to the spin-triplet $P$-wave levels. 
Previous studies have discussed the production of the spin-triplet $D$-wave 
$b\bar{b}$ states \cite{KR,GRdw} and there has been some discussion of 
how one might produce the $1^1P_1 \; c\bar{c}$ state
\cite{suzuki02,kuang02,tuan,ko,CDQ93,QY}.
Up to now, the only observed heavy quarkonium spin-singlet state has been
the $\eta_c(1^1S_0)$, but the Belle Collaboration \cite{Beletac} has just
announced the discovery of the $\eta_c'(2^1S_0)$ in $B$ decays at a mass
of $(3654 \pm 6 \pm 8)$ MeV/$c^2$. 
There have also been a few measurements suggesting the $1^1P_1 
(c\bar{c})$ state in $\bar{p}p$ annihilation experiments 
\cite{r704,e760,e771} but these results have yet to be confirmed.
No $b\bar{b}$ spin-singlet states have yet been seen.

The mass predictions for the singlet states are an important test of
QCD motivated potential models \cite{GI,MR83,LPR92,OS82,MB83,GRR86,IO87,%
GOS84,PJF92,HOOS92} and the applicability of perturbative quantum
chromodynamics to the heavy quarkonia $c\bar{c}$ and $b\bar{b}$ systems
\cite{NPT,PTN86,PT88,FY}, as well as the more recent NRQCD \cite{NRQCD}
approach.  For QCD-motivated potential models the triplet-singlet splittings
test the Lorentz nature of the confining potential with different combinations
of Lorentz scalar, vector, etc., giving rise to different orderings of the
triplet-singlet splittings in the heavy quarkonium P-wave mesons. Furthermore,
the observation of $c\bar{c}$ and $b\bar{b}$ states and the measurement of
their masses is an important validation of lattice QCD calculations
\cite{davies94,davies98,eicker98,bali,manke,okamoto}, which will lead to
greater confidence in their application in extracting electroweak quantities
from hadronic processes.  Under the assumption of a Fermi-Breit Hamiltonian
and only vector-like and scalar-like components in the central potential,
Stubbe and Martin \cite{SM} predicted that the $n^1P_1$ mass lies no lower than
the spin-averaged $^3P_J$ masses (weighted with the factors $2J+1$), denoted by
$n^3P_{\rm cog}$.  Violation of these bounds would indicate a significant
underestimate by \cite{SM} of relativistic effects.

In Table \ref{tab:hfs} we summarize some predictions for hyperfine mass
splittings for P-wave $c\bar{c}$ and $b\bar{b}$ levels. The wide variation in
the predicted splittings demonstrates the need for experimental tests of the
various calculational approaches.

\begin{table}
\caption{Predictions for hyperfine splittings $M(n^3P_{\rm cog}) - 
M(n^1P_1)$ for $c\bar{c}$ and $b \bar b$ levels.
\label{tab:hfs}}
\begin{center}
\begin{tabular}{l c c c c} \hline \hline
Reference & Approach & $n=1$ $c\bar{c}$ & $n=1$ $b\bar{b}$  & $n=2$ $b\bar{b}$
\\ & & (MeV) & (MeV) & (MeV) \\  \hline
GI85 \cite{GI}	   & a &  8 & 2 & 2 \\
MR83 \cite{MR83}   & b &  0 & 0 & 1 \\
LPR92 \cite{LPR92} & c &  4 & 2 & 1 \\
OS82 \cite{OS82}   & d & 10 & 3 & 3 \\
MB83 \cite{MB83}   & e & $-5$ & $-2$ & $-2$ \\
GRR86 \cite{GRR86} & f & $-2$ & $-1$ & $-1$ \\
IO87 \cite{IO87}   & g & $24.1\pm 2.5$ & $3.73 \pm 0.1$ & $3.51 \pm 0.02$ \\
GOS84 $\eta_s=1$ \cite{GOS84} & h & 6 & 3 & 2 \\
GOS84 $\eta_s=0$ \cite{GOS84} & h & 17 & 8 & 6 \\
PJF92 \cite{PJF92} & i & $-20.3 \pm 3.7$ & $-2.5\pm 1.6$ & $ -3.7\pm 0.8$ \\
HOOS92 \cite{HOOS92} & j & $-0.7\pm 0.2$ & $-0.18\pm 0.03$ & $-0.15 \pm 0.03$ \\
PTN86 \cite{PTN86} & j & $-3.6$ & $-0.4$ & $-0.3$ \\
PT88 \cite{PT88} & j & $-1.4$ & $-0.5$ & $-0.4$ \\
SESAM98 \cite{eicker98} & k & -- & $\sim -1$ & --  \\
CP-PACS00 \cite{manke} & l & 1.7--4.0 & 1.6--5.0 & -- \\
\hline \hline
\end{tabular}
\end{center}
\leftline{\qquad $^a$ Potential model with smeared short range 
hyperfine interaction. }
\leftline{\qquad (The splittings are based on masses 
rounded to 1 MeV, not the results}
\leftline{\qquad rounded to 10 MeV as given in Ref.\ \cite{GI}.)}
\leftline{\qquad $^b$ Potential model with long range 
longitudinal color electric field.}
\leftline{\qquad $^c$ Potential model with PQCD corrections to short 
distance piece.}
\leftline{\qquad $^d$ Potential model with smeared hyperfine interaction.}
\leftline{\qquad $^e$ Potential model with smeared hyperfine 
interaction and relativistic corrections}
\leftline{\qquad $^f$ Potential model includes 1-loop QCD corrections.}
\leftline{\qquad $^g$ Potential model with short distance from 2-loop 
PQCD calculation.}  
\leftline{\qquad Results shown for $\Lambda_{\bar{MS}}=200$~MeV.}
\leftline{\qquad $^h$ Potential model with confining potential with 
both Lorentz scalar and vector.}  
\leftline{\qquad $\eta_s$ gives the fraction of the 
confining potential that is pure Lorentz scalar}
\leftline{\qquad versus Lorentz vector.}
\leftline{\qquad $^i$ Potential model. Solution is for Richardson potential and 
$m_c=1.49\pm 0.1$~GeV.}
\leftline{\qquad Other solutions given in Ref.\ \cite{PJF92} are 
consistent with this result within errors.}
\leftline{\qquad $^j$ PQCD}
\leftline{\qquad $^k$ Unquenched nonrelativistic lattice QCD.}
\leftline{\qquad $^l$ Lattice QCD; the result is dependent on the 
value used for $\beta$ and $m_Q$.}
\end{table}
\bigskip

There are two possibilities for producing spin-singlet states.  In the 
first, the system undergoes a magnetic dipole (M1) transition from 
a spin-triplet state to a spin-singlet state.  The predictions 
for $M1$ transitions from the $\ups(n^3S_1)$ levels to the 
$\eta_b (n'^1S_0)$ states, for both favored $M1$ transitions and 
hindered $M1$ transitions with changes of the principal quantum 
number, have been reviewed in Ref. \cite{GRetab}.  The second route
begins with a hadronic transition, from a $n^3S_1$ state to a 
$^1P_1$ state, emitting one or more pions, followed by the electromagnetic
decay of the $^1P_1$ state.

In this paper we examine the production of the spin-singlet $P$-wave $c\bar{c}$
and $b\bar{b}$ states.  We examined the decay chains that start with the $M1$
transition from the $\psi'$ to the $\eta_c'$ in the $c\bar{c}$ system and from
the $\ups(3S)$ to either the $\eta_b(3S)$ or $\eta_b(2S)$ state in the
$b\bar{b}$ system.  In both cases the $M1$ transition is followed by an $E1$
transition to the spin-singlet $2P$ or $1P$ state.  This is in turn followed by
a second $E1$ transition to a $n^1S_0$ state.  In addition, the $2^1P_1$ $b
\bar{b}$ state can undergo an $E1$ transition to the $1^1D_2$ state.  However,
with the current CLEO data set, the only decay chain which has any hope of
being seen in the $b\bar{b}$ system is $\ups(3S)\to \eta_b(2S)\gamma \to h_b
(1P) \gamma\gamma$. We therefore only  present results relevant to this set of
decays.

The decay chains originating with the hadronic transitions are more promising.
We therefore include estimates of branching ratios for chains originating with
the direct hadronic transition $\ups(3S) \to h_b(^1P_1) + \pi^0$ discussed
by Voloshin \cite{voloshin} followed by the radiative decay $h_b(^1P_1)\to
\eta_b(1S) +\gamma$ and the analogous transitions in the charmonium system
$\psi'(2S) \to h_c(^1P_1) + \pi^0 \to \eta_c(1S) \gamma \pi^0$.  Kuang  and Yan
\cite{ky} have also considered the related spin-flip transition $\ups(3S)
\to h_b(^1P_1)+\pi\pi$ which may provide an additional path to the $h_b$.

Searches for the $^1P_1$ states have taken on renewed interest because of the
current data-taking runs of the CLEO Collaboration at the Cornell Electron 
Storage Ring (CESR), which are expected to significantly increase their 
sample of data at the $\ups(3S)$ resonance, and the proposed 
CLEO-c project which will study physics in the charmonium system.

We begin with the $b\bar{b}$ mesons and decay chains involving only radiative
transitions.  To estimate the number of events expected from these decay chains
we need to estimate the radiative partial decay widths between states and the
hadronic partial widths of the appropriate $^1S_0$ and $^1P_1$ states.

The $M1$ transitions from the $\ups(3S)$ to the $\eta_b(3S)$ and
$\eta_b(2S)$ were studied in detail in Ref.\ \cite{GRetab}, which we will use
in what follows.  The $E1$ transitions are straightforward to work out
\cite{KR} and in the nonrelativistic limit are given by
\begin{equation} \label{eqn:stop}
\Gamma(^1S_0 \to {^1P_1} + \gamma) = \frac{4}{3} \alpha \; e_Q^2 \; 
\omega^3 \; |\langle ^1P_1 | r | ^1S_0 \rangle |^2
\end{equation}
\begin{equation} \label{eqn:ptos}
\Gamma(^1P_1 \to {^1S_0} + \gamma) = \frac{4}{9} \alpha \; e_Q^2 \; 
\omega^3 \; |\langle ^1S_0 | r | ^1P_1 \rangle |^2
\end{equation}
\begin{equation} \label{eqn:ptod}
\Gamma(^1P_1 \to ^1D_2 + \gamma) = \frac{8}{9} \alpha \; e_Q^2 \; 
\omega^3 \; |\langle ^1D_2 | r | ^1P_1 \rangle |^2
\end{equation}
where $\alpha = 1/137.036$ is the fine-structure constant, $e_Q$ is the
quark charge in units of $|e|$ ($-1/3$ for $Q=b$), and $\omega$ is the 
photon's energy.  The photon energies, overlap integrals, and partial widths
for the $E1$ transitions between $^1P_1$ and $^1S_0$ levels are given in Table
\ref{tab:E1} and summarized in Fig.\ 1, along with the relevant $M1$
transitions.  The $n^1S_0$ masses were obtained by subtracting the predictions
of Ref.\ \cite{GI} for the $n^3S_1 - n^1S_0$ splittings from the measured
$n^3S_1$ masses, while the $n^1P_1$ masses were obtained by subtracting the
predictions for the $n^3P_{\rm cog}-n^1P_1$ splittings of Ref.\ \cite{GI} from
measured $n^3P_{\rm cog}$ values.  The overlap integrals, 
$\langle r \rangle \equiv \langle ^1L'_{L'} | r | ^1L_L \rangle$,
were evaluated using the wavefunctions of Ref.\ \cite{GI}.  We found 
that the relativistic effects considered in
Ref. \cite{GI} reduce the partial widths by a few percent at most.
Somewhat larger matrix elements were obtained in an inverse-scattering
approach \cite{KR}, except for the highly suppressed $3S \to 1P$ transition,
whose matrix element is very sensitive to details of wave functions
\cite{GR}.

\begin{figure}
\centerline{\epsfysize = 7in \epsffile{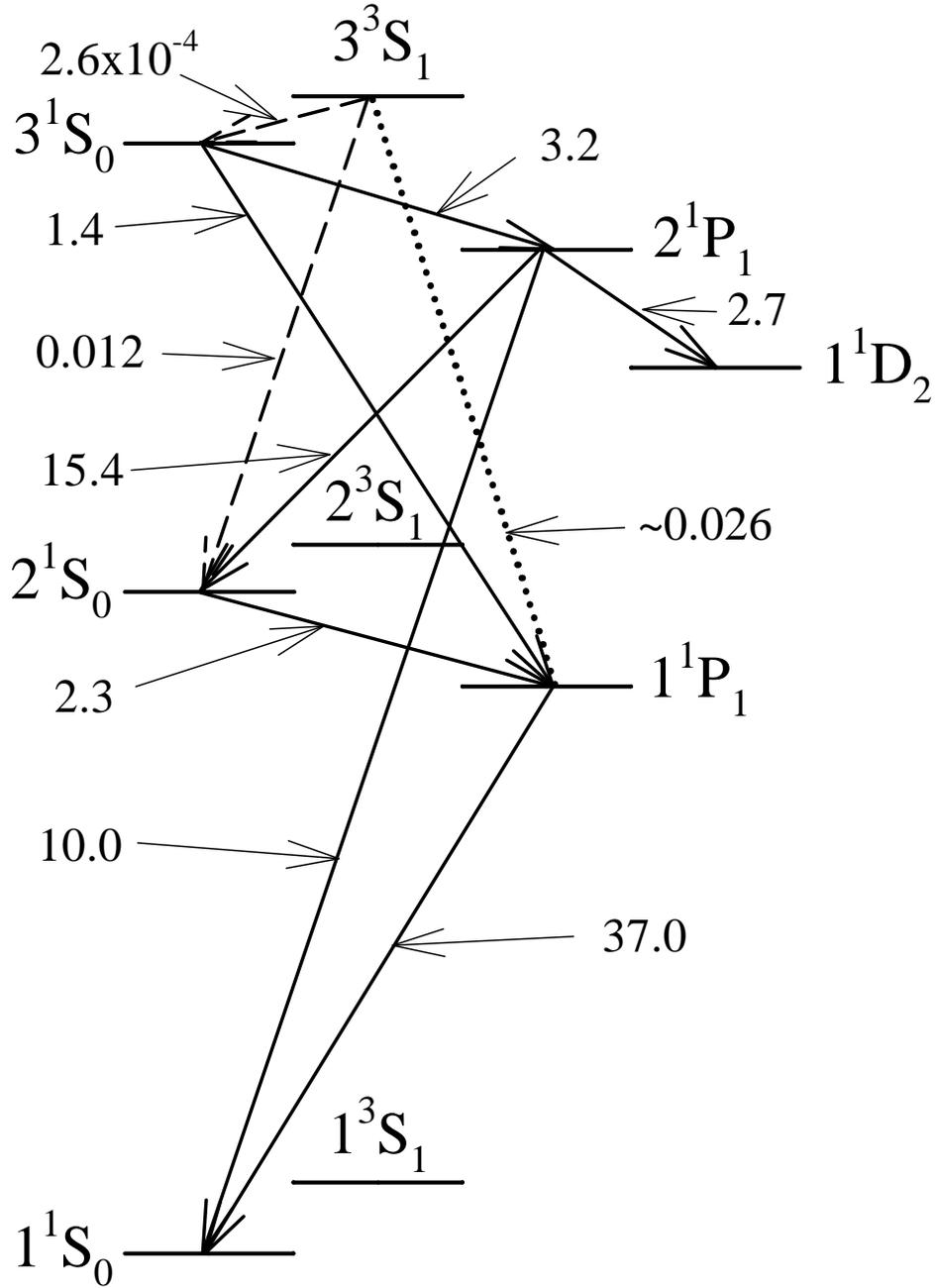}}
\caption{Radiative transitions in the $b\bar{b}$ system.  The dashed 
lines represent $M1$ transitions, the solid lines $E1$ transitions and 
the dotted lines single $\pi^0$ emission.  The transitions are labelled with 
their partial widths given in keV.
\label{fig:m1-trans}}
\end{figure}

\begin{table} \label{tab:e1}
\caption{Radiative electric dipole transitions involving $h_b(1^1P_1)$ and
$h'_b(2^1P_1)$ $b \bar b$ states.
The details of the calculation are given in the text.
\label{tab:E1}}
\begin{center}
\begin{tabular}{l c c c c c} \hline \hline
Transition & $M_i$ & $M_f$ & $\omega$ & $\langle r \rangle$ & $\Gamma$  \\
	& (MeV) & (MeV) & (MeV) & (GeV$^{-1}$) & (keV)  \\ \hline
$3^1S_0 \to 2^1P_1 $ & 10337 & 10258 & 78.7 & $-2.46$ & 3.2  \\
$3^1S_0 \to 1^1P_1 $ & 10337 & 9898 & 430 & 0.126  & 1.4   \\
$2^1P_1 \to 1^1D_2 $ & 10258 & 10148 & 109 & $-1.69$ & 2.7  \\
$2^1P_1 \to 2^1S_0 $ & 10258 & 9996 & 259 &  1.57  & 15.4  \\
$2^1P_1 \to 1^1S_0 $ & 10258 & 9397 & 825 & 0.222  & 10.0  \\
$2^1S_0 \to 1^1P_1 $ & 9996 & 9898 & 97.5 & $-1.53$ & 2.3  \\
$1^1P_1 \to 1^1S_0 $ & 9898 & 9397 & 488 &  0.940 & 37  \\
\hline \hline
\end{tabular}
\end{center}
\end{table}

To estimate the number of events in a particular decay chain requires branching
fractions which depend on knowing all important partial decay widths.  
Inclusive strong decays to gluon and quark final states generally make large
contributions to the total width and have been studied extensively 
\cite{stogg,bgr,ss-gg,chanowitz,KMRR,novikov}.  The relevant theoretical
expressions, including leading-order QCD corrections \cite{bgr}, are
summarized in Ref.\ \cite{KMRR}: 
\begin{equation}
\Gamma(^1S_0\to gg)= {{8\pi\alpha_s^2}\over{3m_Q^2}} |\psi(0)|^2 
\end{equation}
with a multiplicative correction factor of $(1+4.4 {{\alpha_s}\over{\pi}})$ for
$b\bar{b}$ and $(1+4.8 {{\alpha_s}\over{\pi}} )$ for $c\bar{c}$,
\begin{equation} \label{eqn:sp1}
\Gamma(^1P_1 \to ggg) = \frac{20\alpha_s^3}{9\pi m_Q^4} |R_P'(0)|^2 \ln 
(m_Q\langle r \rangle )
\end{equation}
and
\begin{equation} \label{eqn:sp1g}
\Gamma(^1P_1 \to gg+\gamma ) = {{36}\over 5} e_q^2 {\alpha \over 
\alpha_s } \Gamma(^1P_1 \to ggg)
\end{equation}
where we also include the decay $^1P_1\to gg + \gamma$.  

Considerable uncertainties arise in these expressions from the model-dependence
of the wavefunctions and possible relativistic contributions \cite{GI}.  In
addition, the logarithm in the decay $\Gamma(^1P_1 \to ggg) $ is a measure of
the virtuality of the quark emitting the gluon.  Different choices have been
proposed for its argument, introducing further uncertainty.
Rather than evaluating these expressions in a specific potential model,
we can obtain less model-dependent estimates of strong decays by relating 
ratios of theoretical predictions, in which much of the theoretical 
uncertainties factor out, to experimentally measured widths.  Although we
expect the wavefunction at the origin to be slightly larger for the singlet
state than the triplet state, we expect this difference to be much smaller than
the uncertainties mentioned above.

To make our estimates, we will need in addition to Eqs.\ (3)--(5), the
following expressions \cite{KMRR}:
\begin{equation}
\Gamma(^3S_1\to ggg)= 
{{40(\pi^2-9)\alpha_s^3}\over{81m_Q^2}} |\psi(0)|^2 
\end{equation}
with a multiplicative correction factor of $(1-4.9 {{\alpha_s}\over{\pi}})$
for $b\bar{b}$ and $(1-3.7 {{\alpha_s}\over{\pi}} )$ for $c\bar{c}$,
\begin{equation}
\Gamma(^3S_1\to\gamma + gg)= 
{{32(\pi^2-9) e_Q^2 \alpha \alpha_s^2}\over{9m_Q^2}} |\psi(0)|^2 
\end{equation}
with a multiplicative correction factor of $(1-7.4 {{\alpha_s}\over{\pi}})$
for $b\bar{b}$ and $(1-6.7 {{\alpha_s}\over{\pi}} )$ for $c\bar{c}$,
and
\begin{equation} \label{eqn:tp1}
\Gamma(^3P_1 \to q\bar{q}+g) = \frac{8\alpha_s^3 n_f}{9\pi m_Q^4} |R_P'(0)|^2
\ln (m_Q\langle r \rangle )~,
\end{equation}
where the QCD correction factor for the last expression is not known.  Taking
account of decays of the $2^3S_1$ and $3^3S_1$ states to $\pi \pi
\ups(nS)$, lepton pairs, and $\chi_b(nP) \gamma$, as quoted in Ref.\
\cite{PDG}, we find total branching ratios to non-glue final states, and
assume glue to constitute the remainder.  Using branching ratios and total
widths quoted in Ref.\ \cite{PDG}, we then arrive at the estimates summarized
in Table \ref{tab:nSglue} for $\Gamma[\ups(nS) \to {\rm glue}] \equiv \Gamma
[\ups(nS) \to {\rm hadrons}] + \Gamma[\ups(nS) \to \gamma + {\rm hadrons}]$.

\begin{table}
\begin{center}
\caption{Ingredients in estimates of $\Gamma[\ups(nS) \to {\rm glue}]$.
\label{tab:nSglue}}
\begin{tabular}{c c c c c} \hline \hline
$\ups(nS)$ & ${\cal B}(\rm{non-glue})$ & ${\cal B}(\rm{glue})$ & $\Gamma({\rm
tot})$ & $\Gamma({\rm glue})$ \\
state      & (\%) & (\%) & (keV) & (keV) \\ \hline
$\ups(2S)$   & $49.1 \pm 1.5$ & $50.9 \pm 1.5$ & $44 \pm 7$ & $22.4 \pm 3.6$ \\
$\ups(3S)$   & $44.9 \pm 1.4$ & $55.1 \pm 1.4$ & $26.3 \pm 3.5$ & $14.5 \pm
2.0$ \\ \hline \hline
\end{tabular}
\end{center}
\end{table}

Using Eqs.\ (7) and (8), $\alpha_s(\ups _{2S}) = 0.181$, and $\alpha_s(\ups
_{3S})=0.180$ we find $\Gamma[\ups(2S) \to {\rm hadrons})] = 21.8 \pm
3.5$~keV and $\Gamma[\ups(3S) \to {\rm hadrons})] = 14.1\pm 2.0$~keV.  The
ratio of the widths from Eqs.\ (4) and (7):
\begin{equation}
\Gamma(^1S_0\to gg)={{27\pi}\over{5(\pi^2-9)}} {1\over{\alpha_s}}
{{(1+4.4 {{\alpha_s}\over{\pi}} )}\over{(1-4.9{{\alpha_s}\over{\pi}} )}}
\times \Gamma(^3S_1\to ggg)
\end{equation}
results in  $\Gamma[\eta_b(2S) \to {\rm hadrons})] = 4.1 \pm 0.7$~MeV
and $\Gamma[\eta_b(3S) \to {\rm hadrons})] = 2.7 \pm 0.4$~MeV.  

We follow the same procedure to estimate the hadronic width for the $^1P_1$
states, although in this case we need to make use of a theoretical estimate 
for the partial width $^3P_1 \to {^3S_1} \gamma$.  Here we have \cite{KMRR}
\begin{equation}
\Gamma(^1P_1\to {\rm hadrons})={5\over{2n_f}} \times \Gamma(^3P_1 \to {\rm
hadrons})
\end{equation}
where $n_f$ is the number of light quark flavours in the final state 
which we will take to be 3,
ignoring the kinematically suppressed charm-anticharm channel.  This results
in a conservative upper limit for $\Gamma(^1P_1\to {\rm hadrons})$ and hence
a lower limit for the branching ratio of this state to $\gamma + \eta_b$.
As mentioned, the QCD corrections to these widths are not known.  The large
uncertainties arising from the wavefunction and logarithms in Eqs.\ (5) and
(9) cancel out.

The only branching ratios quoted in Ref.\ \cite{PDG} for the $n^3P_1$ states
are for decays to $n'{^3S_1} \gamma$.  Using quark model predictions for the 
radiative transitions and assuming that hadronic decays dominate the 
remainder of the total widths, we can estimate the hadronic partial 
widths of these states.  The results are summarized in Table \ref{tab:strong2}.

\begin{table}
\caption{Partial widths of $^3P_1$ and $^1P_1$ states.
The details of the calculation are given in the text. 
\label{tab:strong2}}
\begin{center}
\begin{tabular}{c c c c c} \hline \hline
$^3P_1$ & $\sum_n{\cal B}(^3P_1 \to n^3S_1 \gamma)^a$ 
	& $\Gamma(^3P_1\to{^3S_1}\gamma)$ & \multicolumn{2}{c}
{$\Gamma(\to {\rm hadrons})$} \\
state & (\%) & (keV) & $^3P_1$ (keV) & $^1P_1$ (keV) \\
\hline
$1^3P_1(b\bar{b})$ & $35 \pm 8$ & 32.8$^b$ & 60.9 & 50.8 \\
		& 	& 28.9$^c$ & 53.7 & 44.7 \\
$2^3P_1(b\bar{b})$ & $29.5 \pm 4.2$ & 25.2$^b$ & 60.2 & 50.2 \\
		& 	& 16.8$^c$ & 40.1 & 33.4 \\
\hline \hline
\end{tabular}
\end{center}
$^a$ Particle Data Group \cite{PDG}

$^b$ Kwong and Rosner \cite{KR}

$^c$ Godfrey and Isgur \cite{GI}
\end{table}

The branching ratios obtained by combining the partial widths given in Tables
\ref{tab:E1} and \ref{tab:strong2} are summarized in Table \ref{tab:br}.

\begin{table}
\caption{Partial widths and branching ratios for spin-singlet $b \bar b$
states.  The details of the calculation are given in the text.
\label{tab:br}}
\begin{center}
\begin{tabular}{l r r r } \hline \hline
Initial         & Final            & Width & ${\cal B}$ \\
 state          & state            & (keV) & (\%) \\ \hline
$3^1S_0$	& $2^1P_1 \gamma $ & 3.2 & 0.12 \\
		& $1^1P_1 \gamma $ & 1.4 & 0.05  \\
		& $gg$ 		& 2700	 & 99.8 \\
$2^1P_1$	& $2^1S_0 \gamma $ &  15.4 & 19.3 \\
		& $1^1S_0 \gamma $ &  10.0 & 12.5 \\
		& $1^1D_2 \gamma $ &  2.7 & 3.4 \\
		& $ggg$		   &  50.2$^a$ & 62.8 \\
		& $\gamma gg$	   &   1.6 & 2.0 \\
$2^1S_0$	& $1^1P_1 \gamma $ &  2.3 & 0.057 \\
		& $gg$		   & 4100  & 99.9 \\
$1^1P_1$	& $1^1S_0 \gamma $ &  37.0 & 41.4 \\
		& $ggg$		   & 50.8$^a$ & 56.8 \\
		& $\gamma gg$	   & 1.6   & 1.8 \\
\hline \hline
\end{tabular}
\end{center}
\leftline{\qquad $^a$ Based on the partial width for $^3P_1 \to {^3S_1}
\gamma$ of Ref.\ \cite{KR} in Table \ref{tab:strong2}.}
\end{table}

To study the singlet $P$-wave $b\bar{b}$ states we considered the two-photon
inclusive transitions $3^3S_1 \gto 3^1S_0 \gto 2^1P_1$ or  $\gto 1^1P_1$ 
and $3^3S_1 \gto 2^1S_0 \gto 1^1P_1$.  In all cases the $^1P_1$ states can
undergo further $E1$ radiative transitions to $^1S_0$ states.  It may be that
this last photon provides a useful tag to distinguish the cascade of interest
from other possible decays involving triplet $P$ and $D$-wave $b\bar{b}$
states.  We use the branching ratios predicted in Ref.\ \cite{GRetab}
for the initial $M1$ transitions, ${\cal B} (\Upsilon(3S) \to \eta_b(3S) +
\gamma) = 0.10 \times 10^{-4}$ and ${\cal B} (\Upsilon(3S) \to \eta_b(2S) +
\gamma) = 4.7\times 10^{-4}$, which correspond to the GI mass-splittings and
wavefunctions \cite{GI} and where the latter result takes into account
relativistic corrections.  Combined with the branching ratios for the
subsequent $E1$ transitions given in Table V the only decay chain that might
yield enough events to be observed is $3^3S_1 \gto 2^1S_0 \gto 1^1P_1$ which
yields roughly 0.3 events per million $\Upsilon(3S)$ states produced.

A more promising approach is the decay chain $\Upsilon(3S)\to {^1P_1} \pi^0$
followed by the $E1$ radiative transition $^1P_1 \to {^1S_0} \gamma$.  Voloshin
estimates ${\cal B} (\Upsilon(3S) \to 1^1P_1 +\pi^0) =0.10 \times 10^{-2}$ 
\cite{voloshin}.  Thus, ${\cal B}[\Upsilon(3S) \to 1^1P_1 +\pi^0 \to 1^1S_0
\gamma] \simeq 4 \times 10^{-4}$, which would yield $\simeq 400$ events per
million $\Upsilon(3S)$ produced.
This signature should be easily seen by the CLEO detector, which has excellent
photon detection capabilities.  Since the recoil of the $1^1P_1$ state is
relatively small, the $488$~MeV photon from the $1^1P_1\to 1^1S_0$ decay
(suitably Doppler-shifted by up to $\pm$ 20 MeV) should provide a useful tag.
Kuang and Yan predict \cite{ky} the partial width for the hadronic transition 
$\Upsilon(3S)\to h_b (1^1P_1) + \pi \pi$ to be 0.1--0.2~keV, giving a
branching ratio of $\sim (3.8$--$7.6)\times 10^{-3}$.  This is substantially 
higher than the value for ${\cal B} (\Upsilon(3S) \to 1^1P_1 +\pi^0)$ 
quoted above so it could provide an alternative path to the $h_b$.  However,
Voloshin \cite{voloshin} does not obtain such a favorable branching ratio for
this process, finding instead $< 10^{-4}$.

We now turn to the charmonium system.  The search for the $h_c$ was discussed
recently by Kuang \cite{kuang02} so we will be brief in our analysis, 
emphasizing aspects that are different from Kuang \cite{kuang02}.
As in the case of $b\bar{b}$ there are two routes to the $h_c$.  The 
first is the decay chain $\psi' \to \eta_c' \gamma \to h_c \gamma$ and 
the second is through the hadronic transition $\psi' \to h_c \pi^0$.

For the first case we need the various radiative widths.  The expression for
the $E1$ width is given by Eq.\ (1), while the rates for magnetic dipole
transitions are given in the nonrelativistic approximation by 
\begin{equation}
\Gamma(^3S_1 \to {^1S_0} + \gamma) = {{4 \alpha e_Q^2}\over{ 3 m_Q^2}}
\omega^3 |\langle f | j_0 (kr/2) | i \rangle |^2 
\end{equation}
\begin{equation}
\Gamma(^1S_0 \to {^3S_1} + \gamma) = {{4 \alpha e_Q^2}\over{ m_Q^2}}
\omega^3 |\langle f | j_0 (kr/2) | i \rangle |^2 
\end{equation}
where we take $m_c=1.628$~GeV.  The results, using the wavefunctions
and $1^1 P_1$ mass of Ref.\ \cite{GI}, are summarized in Table \ref{tab:ccbr}.
To calculate $\Gamma(\psi' \to \eta_c' \gamma)$, we
took $M(\eta_c') = 3654$~MeV, the central value quoted in Ref.\
\cite{Beletac}.  Note that the widths for the hindered $M1$ transitions
are very sensitive to the wave functions.  The hadronic widths for the
$\eta_c'$ and $h_c$  given in Table \ref{tab:ccbr} were obtained using the  
same procedure used for the $b\bar{b}$ hadronic widths:  We relate theoretical
expressions for ratios of the widths to a known measured width and take
$\alpha_s(\psi')=0.236$.  (In contrast to the $b \bar b$ system, the total
width of the $1^3P_1$ $c \bar c$ meson is known \cite{PDG}:  $\Gamma_{\rm tot}
(\chi_{c1}) = 0.88 \pm 0.14$ MeV.)  The predicted result for $h_c \to
{\rm  hadrons}$ is consistent with the NRQCD result obtained by
Bodwin, Braaten and Lepage \cite{bodwin}.  Combining these results we find that
${\cal B}(\psi' \to \eta_c' \gamma) \times  {\cal B} (\eta_c' \to h_c \gamma)
\sim 10^{-6}$, which would yield a modest number of $h_c$ mesons at best.

\begin{table}
\caption{Partial widths and branching ratios for spin-singlet $c\bar{c}$
states.  In column 5, ${\cal O}$  represents the operator relevant to 
the particular electromagnetic transition; ${\cal O}= r$ (GeV$^{-1}$) 
for $E1$ transitions and ${\cal O}=j_0(kr/2)$ for $M1$ transitions. 
The details of the calculation are given in the text.
\label{tab:ccbr}}
\begin{center}
\begin{tabular}{l c c c c c c c } \hline
Initial         & Final    & $M_i$ & $M_f$ & $\omega$ & $\langle f | {\cal O}
 | i \rangle$   & Width & ${\cal B}$ \\
state           & state  & (MeV)  & (MeV)  & (MeV) &       
					& (keV) & (\%) \\ \hline
$2^3S_1$	& $2^1S_0 \gamma $ & 3686 & 3654 & 31.8 & 0.982 
					&  0.051 & 0.018 \\
		& $1^1S_0 \gamma $ & 3686 & 2980 & 638 & 0.151
					&  9.7  & 3.5  \\
$2^1S_0$	& $1^1P_1 \gamma $ & 3654 & 3517 & 134.4 & $-2.21$ 
					& 51.3 & 0.69 \\
		& $1^3S_1 \gamma$  & 3654 & 3097 & 515 & $-0.0973$ 
					& 6.3  & 0.084 \\
		& $gg$		   &      &      &       & 
					&  7400 & 99.2 \\
$1^1P_1$	& $1^1S_0 \gamma $ & 3517 & 2980 & 496 & 1.42
					& 354  & 37.7 \\
		& $ggg$		   &      &      &     &
					& 533  & 56.8 \\
		& $\gamma gg$	   &      &      &     &
					&  52  & 5.5 \\
\hline
\end{tabular}
\end{center}
\end{table}

As in the case of the $h_b$, a more promising avenue is the single pion
transition $\psi' \to h_c \pi^0$ followed by the radiative transition 
$h_c \to \eta_c \gamma$, where the photon is expected to have an energy 
very close to 496~MeV and can be used to tag the event.  Using the 
branching ratio of ${\cal B} (\psi' \to h_c \pi^0)=0.1 \% $ predicted 
by Voloshin \cite{voloshin}  (see also Ref.\ \cite{ko,ky}) and the branching
ratio given for $h_c \to \eta_c \gamma$ in Table \ref{tab:ccbr}, we obtain 
${\cal B}(\psi' \to h_c \pi^0) \times  {\cal B} (h_c \to \eta_c \gamma) =
3.8 \times 10^{-4}$, which is substantially larger than the decay chain
proceeding only via radiative transitions.  In his recent paper Kuang
\cite{kuang02} finds $h_c$ production to be sensitive to $^3S_1-{^3D_1}$
mixing, so that a measurement of ${\cal B}(\psi' \to h_c \pi^0)$ would be a
useful test of detailed mixing schemes between the $\psi'$ and the $\psi(3770)
\equiv \psi''$, some of which are discussed in Ref.\ \cite{JRmix,GKO}.

Another promising approach for the detection of the $h_c$ has recently 
been proposed by Suzuki \cite{suzuki02}.  He suggests looking for the 
$h_c$ by measuring the final state $\gamma\eta_c$ of the cascade $B\to 
h_c K/K^* \to \gamma \eta_c K/K^*$. This channel is especially timely 
given the announcement by the Belle Collaboration of the discovery of
the $\eta_c'(2^1S_0)$ in $B$ decays \cite{Beletac} and, previously, the 
observation of the related decay, $B\to \chi_0 K$ \cite{belle01}.

In the case of the S-wave ($^1S_0$) states, one should also bear in mind
that $\gamma \gamma$ collisions have been used to observed the $\eta_c$ in
several experiments (see \cite{PDG}).  One candidate for $\gamma \gamma \to
\eta_b$ with mass $9.30 \pm 0.02 \pm 0.02$ GeV/$c^2$ (consistent, however, with
background) has been reported by the ALEPH Collaboration \cite{ALetab}.

To conclude, we have explored different means of looking for the 
$^1P_1$ states in heavy quarkonium.  In both the $b\bar{b}$ and 
$c\bar{c}$ systems the $1^1P_1$ state can be reached via the chain 
$^3S_1 \to {^1S_0} + \gamma \to 1^1P_1 +\gamma\gamma$.  However, in both 
systems one only expects of the order of a few events per million 
$\Upsilon(3S)$ or $\psi'$ produced.  In both systems, a more promising 
avenue is the transition $^3S_1 \to {^1P_1} + \pi^0$ followed by the 
$E1$ radiative transition to the $1^1S_0$ state which would yield 
several hundred events per million $\Upsilon(3S)$ or $\psi'$'s produced.
The alternative suggestion \cite{ky} of searching for the transitions
$^3S_1 \to {^1P_1} + \pi \pi$ also is worth pursuing.

One of us (J.L.R.) thanks Steve Olsen and San Fu Tuan for informative
discussions.  The authors thank Christine Davies for helpful communications.
This work was supported in part by the United
States Department of Energy through Grant No.\ DE FG02 90ER40560
and the Natural Sciences and Engineering Research Council of Canada.

\def \ajp#1#2#3{Am.\ J. Phys.\ {\bf#1}, #2 (#3)}
\def \apny#1#2#3{Ann.\ Phys.\ (N.Y.) {\bf#1}, #2 (#3)}
\def \app#1#2#3{Acta Phys.\ Polonica {\bf#1}, #2 (#3)}
\def \arnps#1#2#3{Ann.\ Rev.\ Nucl.\ Part.\ Sci.\ {\bf#1}, #2 (#3)}
\def \art{and references therein}
\def \cmts#1#2#3{Comments on Nucl.\ Part.\ Phys.\ {\bf#1}, #2 (#3)}
\def \cn{Collaboration}
\def \cp89{{\it CP Violation,} edited by C. Jarlskog (World Scientific,
Singapore, 1989)}
\def \efi{Enrico Fermi Institute Report No.\ }
\def \epjc#1#2#3{Eur.\ Phys.\ J. C {\bf#1}, #2 (#3)}
\def \f79{{\it Proceedings of the 1979 International Symposium on Lepton and
Photon Interactions at High Energies,} Fermilab, August 23-29, 1979, ed. by
T. B. W. Kirk and H. D. I. Abarbanel (Fermi National Accelerator Laboratory,
Batavia, IL, 1979}
\def \hb87{{\it Proceeding of the 1987 International Symposium on Lepton and
Photon Interactions at High Energies,} Hamburg, 1987, ed. by W. Bartel
and R. R\"uckl (Nucl.\ Phys.\ B, Proc.\ Suppl., vol.\ 3) (North-Holland,
Amsterdam, 1988)}
\def \ib{{\it ibid.}~}
\def \ibj#1#2#3{~{\bf#1}, #2 (#3)}
\def \ichep72{{\it Proceedings of the XVI International Conference on High
Energy Physics}, Chicago and Batavia, Illinois, Sept. 6 -- 13, 1972,
edited by J. D. Jackson, A. Roberts, and R. Donaldson (Fermilab, Batavia,
IL, 1972)}
\def \ijmpa#1#2#3{Int.\ J.\ Mod.\ Phys.\ A {\bf#1}, #2 (#3)}
\def \ite{{\it et al.}}
\def \jhep#1#2#3{JHEP {\bf#1}, #2 (#3)}
\def \jpb#1#2#3{J.\ Phys.\ B {\bf#1}, #2 (#3)}
\def \lg{{\it Proceedings of the XIXth International Symposium on
Lepton and Photon Interactions,} Stanford, California, August 9--14 1999,
edited by J. Jaros and M. Peskin (World Scientific, Singapore, 2000)}
\def \lkl87{{\it Selected Topics in Electroweak Interactions} (Proceedings of
the Second Lake Louise Institute on New Frontiers in Particle Physics, 15 --
21 February, 1987), edited by J. M. Cameron \ite~(World Scientific, Singapore,
1987)}
\def \kdvs#1#2#3{{Kong.\ Danske Vid.\ Selsk., Matt-fys.\ Medd.} {\bf #1},
No.\ #2 (#3)}
\def \ky85{{\it Proceedings of the International Symposium on Lepton and
Photon Interactions at High Energy,} Kyoto, Aug.~19-24, 1985, edited by M.
Konuma and K. Takahashi (Kyoto Univ., Kyoto, 1985)}
\def \mpla#1#2#3{Mod.\ Phys.\ Lett.\ A {\bf#1}, #2 (#3)}
\def \nat#1#2#3{Nature {\bf#1}, #2 (#3)}
\def \nc#1#2#3{Nuovo Cim.\ {\bf#1}, #2 (#3)}
\def \nima#1#2#3{Nucl.\ Instr.\ Meth. A {\bf#1}, #2 (#3)}
\def \np#1#2#3{Nucl.\ Phys.\ {\bf#1}, #2 (#3)}
\def \npbps#1#2#3{Nucl.\ Phys.\ B Proc.\ Suppl.\ {\bf#1}, #2 (#3)}
\def \os{XXX International Conference on High Energy Physics, Osaka, Japan,
July 27 -- August 2, 2000}
\def \PDG{Particle Data Group, D. E. Groom \ite, \epjc{15}{1}{2000}}
\def \pisma#1#2#3#4{Pis'ma Zh.\ Eksp.\ Teor.\ Fiz.\ {\bf#1}, #2 (#3) [JETP
Lett.\ {\bf#1}, #4 (#3)]}
\def \pl#1#2#3{Phys.\ Lett.\ {\bf#1}, #2 (#3)}
\def \pla#1#2#3{Phys.\ Lett.\ A {\bf#1}, #2 (#3)}
\def \plb#1#2#3{Phys.\ Lett.\ B {\bf#1}, #2 (#3)}
\def \pr#1#2#3{Phys.\ Rev.\ {\bf#1}, #2 (#3)}
\def \prc#1#2#3{Phys.\ Rev.\ C {\bf#1}, #2 (#3)}
\def \prd#1#2#3{Phys.\ Rev.\ D {\bf#1}, #2 (#3)}
\def \prl#1#2#3{Phys.\ Rev.\ Lett.\ {\bf#1}, #2 (#3)}
\def \prp#1#2#3{Phys.\ Rep.\ {\bf#1}, #2 (#3)}
\def \ptp#1#2#3{Prog.\ Theor.\ Phys.\ {\bf#1}, #2 (#3)}
\def \rmp#1#2#3{Rev.\ Mod.\ Phys.\ {\bf#1}, #2 (#3)}
\def \rp#1{~~~~~\ldots\ldots{\rm rp~}{#1}~~~~~}
\def \rpp#1#2#3{Rep.\ Prog.\ Phys.\ {\bf#1}, #2 (#3)}
\def \sing{{\it Proceedings of the 25th International Conference on High Energy
Physics, Singapore, Aug. 2--8, 1990}, edited by. K. K. Phua and Y. Yamaguchi
(Southeast Asia Physics Association, 1991)}
\def \slc87{{\it Proceedings of the Salt Lake City Meeting} (Division of
Particles and Fields, American Physical Society, Salt Lake City, Utah, 1987),
ed. by C. DeTar and J. S. Ball (World Scientific, Singapore, 1987)}
\def \slac89{{\it Proceedings of the XIVth International Symposium on
Lepton and Photon Interactions,} Stanford, California, 1989, edited by M.
Riordan (World Scientific, Singapore, 1990)}
\def \smass82{{\it Proceedings of the 1982 DPF Summer Study on Elementary
Particle Physics and Future Facilities}, Snowmass, Colorado, edited by R.
Donaldson, R. Gustafson, and F. Paige (World Scientific, Singapore, 1982)}
\def \smass90{{\it Research Directions for the Decade} (Proceedings of the
1990 Summer Study on High Energy Physics, June 25--July 13, Snowmass, Colorado),
edited by E. L. Berger (World Scientific, Singapore, 1992)}
\def \tasi{{\it Testing the Standard Model} (Proceedings of the 1990
Theoretical Advanced Study Institute in Elementary Particle Physics, Boulder,
Colorado, 3--27 June, 1990), edited by M. Cveti\v{c} and P. Langacker
(World Scientific, Singapore, 1991)}
\def \yaf#1#2#3#4{Yad.\ Fiz.\ {\bf#1}, #2 (#3) [Sov.\ J.\ Nucl.\ Phys.\
{\bf #1}, #4 (#3)]}
\def \zhetf#1#2#3#4#5#6{Zh.\ Eksp.\ Teor.\ Fiz.\ {\bf #1}, #2 (#3) [Sov.\
Phys.\ - JETP {\bf #4}, #5 (#6)]}
\def \zpc#1#2#3{Zeit.\ Phys.\ C {\bf#1}, #2 (#3)}
\def \zpd#1#2#3{Zeit.\ Phys.\ D {\bf#1}, #2 (#3)}


\begin{thebibliography}{99}

\bibitem{revs} For reviews see W. Kwong, J. L. Rosner, and C. Quigg,
\arnps{37}{325}{1987}; W. Buchm\"uller and S. Cooper, in {\it High
Energy Electron--Positron Physics,} edited by A. Ali and P. S\"oding,
World Scientific, Singapore, 1988, p.~412; D. Besson and T. Skwarnicki,
\arnps{43}{333}{1993}; E. Eichten and C. Quigg, \prd{49}{5845}{1994}.

\bibitem{KR} W. Kwong and J. L. Rosner, \prd{38}{279}{1988}.

\bibitem{GRdw} S. Godfrey and J. L. Rosner, \prd{64}{097501}{2001}. 

\bibitem{suzuki02} 
M. Suzuki, hep-ph/0204043.

\bibitem{kuang02}
Y. P. Kuang, hep-ph/0201210.

\bibitem{tuan} S. F. Tuan, Commun.\ Theor.\ Phys.\ {\bf 26}, 381 (1996).

\bibitem{ko}
P. Ko, \prd{52}{1710}{1995}.

\bibitem{CDQ93}
K.-T. Chao, Y.-B. Ding, and D.-H. Qin, \plb{301}{282}{1993}.

\bibitem{QY} C.-F. Qiao and C. Yuan, \prd{63}{014007}{2001}.

\bibitem{Beletac}
Belle \cn, K. Abe \ite, presented at FPCP Conference, Philadelphia, PA,
May 16--18, 2002.

\bibitem{r704} C. Baglin \ite, \plb{171}{135}{1986}.

\bibitem{e760} T. A. Armstrong \ite, \prl{69}{2337}{1992}.

\bibitem{e771} L. Antoniazzi \ite, \prd{50}{4258}{1994}.

\bibitem{GI} S. Godfrey and N. Isgur, \prd{32}{189}{1985}.

\bibitem{MR83} P. Moxhay and J. L. Rosner, \prd{28}{1132}{1983}.
 
\bibitem{LPR92}
 D. B. Lichtenberg, E. Predazzi, and R. Rocaglia, \prd{45}{3268}{1992}.

\bibitem{OS82}
S. Ono and F. Sch\"oberl, \plb{118}{419}{1982}.

\bibitem{MB83}
R. McClary and N. Byers, \prd{28}{1692}{1983}.

\bibitem{GRR86}
S. N. Gupta, S. F. Radford and W. W. Repko, \prd{34}{201}{1986}.

\bibitem{IO87}
I. Igi and S. Ono, \prd{36}{1550}{1987}.

\bibitem{GOS84}
H. Grotch, D. A. Owen, and K. J. Sebastian, \prd{30}{1924}{1984}.

\bibitem{PJF92}
P. J. Franzini, \plb{296}{199}{1992}.

\bibitem{HOOS92}
F. Halzen, C. Olson,  M. G. Olsson, and M. L. Stong, \plb{283}{379}{1992}.

\bibitem{NPT} 
Y. J. Ng, J. Pantaleone, and S.-H. H. Tye, \prl{55}{916}{1985}.

\bibitem{PTN86}
J. Pantaleone, Y. J. Ng, and S.-H. H. Tye, \prd{33}{777}{1986}.

\bibitem{PT88}
J. Pantaleone and S.-H. H. Tye, \prd{37}{3337}{1988}.

\bibitem{FY} F. J. Yndur\'ain, Lectures at the XVII International 
School of Physics ``QCD: Perturbative or Nonperturbative'', Lisbon, 
1999, hep-ph/9910399.

\bibitem{NRQCD} G. P. Lepage and B. A. Thacker, \npbps{4}{199}{1988};
B. A. Thacker and G. P. Lepage, \prd{43}{196}{1991}.

\bibitem{davies94} C. T. H. Davies \ite, \prd{50}{6963}{1994}.

\bibitem{davies98} C. T. H. Davies \ite, \prd{58}{054505}{1998}.

\bibitem{eicker98} N. Eicker \ite, \prd{57}{4080}{1998}.

\bibitem{bali} G. S. Bali, K. Schilling, and A. Wachter, \prd{56}{2566}{1997}.

\bibitem{manke} T. Manke \ite, \prd{62}{114508}{2000}.

\bibitem{okamoto}
M. Okamoto \ite, hep-lat/0112020.

\bibitem{SM}
J. Stubbe and A. Martin, \plb{271}{208}{1991}.

\bibitem{GRetab} S. Godfrey and J. L. Rosner, \prd{64}{074011}{2001}. 

\bibitem{voloshin}
M. B. Voloshin, Sov. J. Nucl. Phys. {\bf 43}, 1011 (1986);
M. B. Voloshin and V. I. Zakharov, \prl{45}{688}{1980}.

\bibitem{ky}
Y. P. Kuang and T. M. Yan, \prd{24}{2874}{1981};
T. M. Yan, \prd{22}{1652}{1980};
Y. P. Kuang, S. F. Tuan and T. M. Yan, \prd{37}{1210}{1988}.

\bibitem{GR} A. Grant and J. L. Rosner, \prd{46}{3862}{1992}.

\bibitem{stogg}
T. Appelquist and H. D. Politzer, \prl{34}{43}{1975};
A. De Rujula and S. L. Glashow, \prl{34}{46}{1975}.

\bibitem{bgr}
R. Barbieri, R. Gatto, and E. Remiddi, \plb{61}{465}{1976}.

\bibitem{ss-gg}
R. Barbieri, G. Curci, E. d'Emilio, and E. Remiddi, \np{B154}{535}{1979}.

\bibitem{chanowitz}
M. Chanowitz, \prd{12}{918}{1975}.

\bibitem{KMRR} W. Kwong, P. B. Mackenzie, R. Rosenfeld, and J. L. Rosner,
\prd{37}{3210}{1988}.

\bibitem{novikov}
V. A. Novikov, L. B. Okun, M. A. Shifman, A. I. Vainshtein, M. B. Voloshin, 
and V. I. Zakharov, \prp{41}{1}{1978}.

\bibitem{PDG} Particle Data Group, D. E. Groom \ite, \epjc{15}{1}{2000}.

\bibitem{bodwin} 
G. T. Bodwin, E. Braaten, and G. P. Lepage, \prd{46}{R1914}{1992}. 

\bibitem{JRmix} J. L. Rosner, \prd{64}{094002}{2001}.

\bibitem{GKO}
S. Godfrey, G. Karl, and P. J. O'Donnell, \zpc{31}{77}{1986}.

\bibitem{belle01}
Belle \cn, K. Abe \ite, hep-ex/0107050.

\bibitem{ALetab} ALEPH \cn, A. Heister \ite, \plb{530}{56}{2002}.

\end{thebibliography}
\end{document}